\documentclass[useAMS,usenatbib]{mn2e}
\usepackage{times,txfonts,graphicx,xspace,natbib,color,amssymb, caption, tabularx, aas_macros}
\usepackage[normalem]{ulem}
\usepackage[dvipsnames]{xcolor}
\newcommand{\be}{\begin{eqnarray}}
\newcommand{\en}{\end{eqnarray}}
\newcommand{\pa}{\partial}
\newcommand{\f}{\frac}

\newcommand\bb[1]{\mbox{\boldmath{$#1$}}}
 
\newcommand\bcdot{\bb{\cdot}}
\newcommand\btimes{\bb{\times}}

\title[He fingers in ICM]{On the Helium fingers in the intracluster medium}
\author[Gupta et al.]
{Shubhadeep Sadhukhan$^{1}$\thanks{Email: deep@iitk.ac.in}, Himanshu Gupta$^{1}$\thanks{Email: hiugupta@iitk.ac.in}, and Sagar Chakraborty$^{1}\thanks{Email:sagarc@iitk.ac.in}$
\\
$^{1}$ Department of Physics, Indian Institute of Technology Kanpur, U.P.-208016, India.}

\begin{document}

\maketitle

\label{firstpage}

\begin{abstract}
	In this paper we investigate the convection phenomenon in the intracluster medium (the weakly-collisional magnetized inhomogeneous plasma permeating galaxy clusters) where the concentration gradient of the Helium ions is not ignorable. 	
	To this end, we build upon the general machinery employed to study the salt finger instability found in the oceans. The salt finger instability is a form of double diffusive convection where the diffusions of two physical quantities---heat and salt concentrations---occur with different diffusion rates.
	 The analogous instability in the intracluster medium may result owing to the magnetic field mediated anisotropic diffusions of the heat and the Helium ions (in the sea of the Hydrogen ions and the free electrons). These two diffusions have inherently different diffusion rates. Hence the convection caused by the onset of this instability is an example of double diffusive convection in the astrophysical settings.
	 A consequence of this instability is the formation of the vertical filamentary structures having more concentration of the Helium ions with respect to the immediate neighbourhoods of the filaments. We term these structures as Helium fingers in analogy with the salt fingers found in the case of the salt finger instability. 
	Here we show that the width of a Helium finger scales as one-fourth power of the radius of the inner region of the intracluster medium in the supercritical regime.	We also determine the explicit mathematical expression of the criterion for the onset of the heat-flux-driven buoyancy instability modified by the presence of inhomogeneously distributed Helium ions.
\end{abstract}


\section{Introduction} 
\label{sec:Introduction}
	The intracluster medium (ICM) that permeats galaxy clusters is a weakly collisional, high-beta, multi-species plasma~
	\citep{carilli_araa02, peterson_pr06}. 
	Owing to the presence of the magnetic field, the transports of the heat and the momentum, and the diffusion of ions are anisotropic in the ICM. 
	Straightforward linear stability analysis
	establishes that convection ensues in the ICM---assumed homogeneous and initially at rest---due to the onset of {convective instabilities facilitated by anisotropy introduced in the system by the magnetic field. The
	magnetothermal instability (MTI)~\citep{balbus_apj00, balbus_apj01} and the
	heat-flux-driven buoyancy instability (HBI)~\citep{quataert_apj08} are special cases of such instabilities that appear when the magnetic field is respectively perpendicular and parallel to the temperature gradient (assumed antiparallel to the gravitational field).} 

	Since the ICM is composed of mainly electrons, Hydrogen ions, 
	Helium ions and a very small fraction of heavier elements, it is intuitive, and theoretically proposed, that the different ions would diffuse 
	to enrich~\citep{fabian_mnras77, rephaeli_apj1978, abramopoulos_apj1981, qin_apjl00, chuzhoy_mnras03, chuzhoy_mnras04} the cluster core with 
	heavy elements along with Helium ions. Moreover, since both Hydrogen and Helium in the ICM are fully ionized rendering measurement of their abundances using traditional spectroscopic techniques difficult, there is no full-proof justification that the elements are homogeneously mixed. {The theoretical investigations of convective instabilities in the ICM mostly consider it as a homogeneous medium for simplicity (see~\citet{pessah_apj13, berlok_apj15, berlok_apj16, berlok_apj16b} for exceptions)}. This is rather unsatisfactory. In fact it has been argued that during the diffusion process a nonzero gradient of mean molecular weight is created in the ICM~\citep{qin_apjl00, peng_apj09, shtykovskiy_10, bulbul_aanda11} and it has non-trivial effects on the HBI and the MTI~\citep{pessah_apj13}. Thus, in spite of its simplicity, the assumption of a homogeneous medium must be {relaxed} in order to understand the interplay between the process of Helium sedimentation and the convective instability arising out of inhomogeneity thus created. 

As done in this paper, in the first approximation, one may ignore the far rarer heavier ions and consider 
the plasma as a dilute binary mixture of Hydrogen and Helium ions (and, of course, electrons). 
The nature of Helium sedimentation---a very important phenomenon in itself---on to the cluster 
core may affect {the inference of} different physical quantities such as the total mass of the cluster 
\citep{qin_apjl00}, cosmological parameters from the observational data \citep{markevitch_pr07, peng_apj09}; 
and thus may need to be considered for more precise calculations~\citep{mroczkowski_apjl11, bulbul_aanda11} of the cluster 
mass and the Helium abundance profile in the cluster core as determined from observational X-ray data.

This simultaneous presence of the temperature and the composition gradients is reminiscent of familiar 
earthly situations conducive to the common double diffusive instabilities found in the oceans where 
vertical gradients of temperature and salinity are found naturally. In the gravitationally stably 
stratified case where hot salty water is put on top of cold fresh water, salt-fingers are 
observed~\citep{stern_tellus60} because of the difference in the diffusivities of salt and heat --- salt 
diffuses relatively slower. Stability analyses~\citep{stern_tellus60, schmitt_dsrpor79} establish that the 
salt-fingers are observed in the supercritical regime well beyond the scope of the linear stability analysis. 
More elaborate treatment of the double diffusive convection is considered 
in many papers~\citep{monchick_pof67, mcdougall_pio81a, kunze_jomr87, mcdougall_pio81b, ozgokmen_pof98, radko_jfm08, radko_jfm10}. 
It interests us to note that double diffusive convection has been invoked in the astrophysical 
contexts to investigate convection in steller interiors~\citep{ulrich_apj72}, photospheric composition 
of low mass giants~\citep{charbonnel_aanda07_l15}, and 
galactic evolution of $^3\textrm{He}$~\citep{charbonnel_aanda07_l29}.

In view of the above we envisage, in the context of the ICM, the existence and the importance of 
an instability analogous to the salt finger instability. Therefore, we extend the study of double 
diffusive convection to the situations of astrophysical interest in which the plasma
is only weakly-collisional (meaning that the mean free path for particles to 
interact is much larger than the Larmor radius) and in which
transport properties are anisotropic with respect to the direction of 
the magnetic field. In the inner region of the galaxy cluster where the temperature increases with the distance from 
the core of the galaxy cluster~\citep{vikhlinin_apjl06}, the mean molecular 
weight (or equivalently concentration of Helium) is also predicted~\citep{bulbul_aanda11} to have a positive gradient. Consequently, we expect formation of fingers 
(analogous to the salt fingers in some geophysical contexts) in the supercritical regime of the convective instability.
We would call such columnar structures, effecting radial mixing of the ICM, Helium fingers 
for obvious reasons: in these fingers concentration of Helium is relatively higher. 
In this paper, we investigate how the width of the Helium fingers scale with the system parameters and 
find for it a power law behaviour w.r.t. the gradients of the temperature and the mean molecular weight.
\section{Framework for Linear Stability Analysis} 
\label{sec:RBA}
In the present work, we have exclusively confined ourselves to studying the aforementioned double diffusive convection in an idealized setup analogous to parallel-plate-Rayleigh-B\'enard-convection geometry because it conveniently brings forth the essential physics behind the phenomenon under study. Owing to complex nonlinear character of the fluid system, in general, the problem of convection in arbitrary geometry is not always analytically tractable. 

Consider two horizontal parallel plates of infinite extent. The bottom and the top plates are maintained at constant temperatures $T=T_{\rm bottom}$ and $T=T_{\rm
top}$ respectively; and also at constant concentrations (of, say, Helium ions) $c=c_{\rm bottom}$ and $c=c_{\rm top}$, respectively. A weakly-collisional dilute plasma that is a binary mixture of the Hydrogen and the Helium ions (along with free the electrons) is there at rest (fluid velocity $\bb{u}=0$) between
the plates. The mass density of the fluid is $\rho$. The acceleration due to gravity $\bb{g}=-g\hat{\bb{z}}$ is acting
vertically downwards.   Additionally, let's assume that there exists an external  uniform magnetic field $\bb{B}$ lying in the vertical $x$-$z$ plane, i.e., $\bb{B}=B\bb{\hat{b}}=B(\sin\phi\bb{\hat{x}}+\cos\phi\bb{\hat{z}})$. Here $\phi$ is angle between the uniform magnetic field and $z$-axis, and $\hat{\bb{b}}\equiv \bb{B}/B = (b_x,0, b_z)$. Also, it is of use for latter convenience to recall that the plasma-beta parameter is defined as $\beta\equiv{P}/({B^2/4\pi})$, where $P$ is the isotropic plasma pressure.

Hence, in view of the above, at the initial equilibrium state $(\mathbf{u_{\rm eq}},\rho_{\rm eq},P_{\rm eq}, T_{\rm eq},c_{\rm eq},\mathbf{B_{\rm eq}})$ of the fluid system in hand, following relations hold:
\be  
&& \mathbf{u}_{\rm eq}=0\,,\\
&&\f{dP_{\rm eq}}{dz}=-\rho_{\rm eq} g\,,\\
&&T_{\rm eq}= T_{\rm bottom}-\Delta T\,\left(\f{z}{d}\right)\,,\label{eq:T*}\\
&&c_{\rm eq}=c_{\rm bottom}-\Delta c\, \left(\f{z}{d}\right)\,,\label{eq:c*}\\
&&\mathbf{B_{\rm eq}}=B_{\rm eq}\mathbf{\hat{b}}=B_{\rm eq}(\sin\phi\mathbf{\hat{x}}+\cos\phi\mathbf{\hat{z}})\,.
\en
Here $\Delta T \equiv T_{\rm bottom} - T_{\rm top}$, $\Delta c \equiv c_{\rm bottom} - c_{\rm top}$, and  $d$ is the vertical separation between the two plates.

As far as the governing equations of motion of this system is concerned, they are the following well-known set of equations (see also~\citet{pessah_apj13,himanshu_pla16}): 
\be &&\f{\pa \rho}{\pa t}+\bb{\nabla}\bcdot(\rho \mathbf{u})=0 \,,
\label{eq:rho}\\
&&\frac{\pa\mathbf{u}}{\pa t} + \mathbf{u} \bcdot \mathbf{\nabla}\mathbf{u} =-\f{1}{\rho}\bb{\nabla} \bcdot \left({\sf{P}} + 
\f{\bb{B}^2}{8\pi}{\sf{I}} - \f{{B^2}}{4\pi}\hat{\mathbf{b}}\hat{\mathbf{b}}\right) + \mathbf{g} \,, \,\,\,\,
\label{eq:v}\\
&&\f{\pa \mathbf{B}}{\pa t}=\bb{\nabla}\btimes(\mathbf{u}\btimes\mathbf{B})+ \eta \nabla^2\mathbf{B}\,,
\label{eq:b2}\\
&& \rho T\left(\frac{\pa s}{\pa t} + \mathbf{u} \bcdot \mathbf{\nabla}s \right)=
(p_\bot-p_\parallel)\f{d}{dt}\ln\f{B}{\rho^{\gamma-1}} 
+ \bb{\nabla} \bcdot \left[\mathbf{\chi\mathbf{\hat{b}}(\mathbf{\hat{b}}\bcdot
\bb{\nabla})T}\right]
\,,
\label{eq:S}\\ 
&&\frac{\pa c}{\pa t} + \mathbf{u} \bcdot \mathbf{\nabla}c = \bb{\nabla} \bcdot \left[\mathbf{D\mathbf{\hat{b}}(\mathbf{\hat{b}}\bcdot
\bb{\nabla})}c\right]
\,.
\label{eq:C}
\en
In these equations  the specific entropy, the adiabatic index, and the electrical resistivity (or the magnetic diffusivity)  have respectively been denoted by $s$, $\gamma$, and $\eta$. Also, $\chi \approx 6 \times 10^{-7} T^{5/2} \,
\textrm{erg cm$^{-1}$ s$^{-1}$ K$^{-1}$}$~\citep{spitzer_book62,
	braginskii_rpp65} is the thermal conductivity and $D$ is the coefficient of the particle diffusion. For the later usage, we define coefficient of thermal diffusion 
$\lambda\equiv \chi/\rho c_p$, where $c_p\equiv T({\partial s}/{\partial T})_P$ is 
the heat capacity at constant pressure. Pressure tensor ${\sf{P}}$ may be expanded~\citep{hollweg_jgr85} as
	$p_\bot {\sf{I}} + (p_\parallel - p_\bot) \hat{\mathbf{b}} 
	\hat{\mathbf{b}}$,
	where
	$\sf{I}$ is the identity matrix. 
	Subscripts $\bot$ and $\parallel$  symbolize the directions 
	perpendicular and parallel to $\hat{\mathbf{b}}$. It may be noted that the background temperature (and concentration) must at most be linearly dependent (see equations~(\ref{eq:T*}) and (\ref{eq:c*})) on the distance along the direction of the magnetic field in order to ensure a well-defined steady background flux of heat or particles. It has, thus, been implicitly assumed that the coefficients $\chi$ and $D$ are constants that is justified under the local approximation that is used in the paper. This assumption is relaxed in, e.g.,~ \citet{latter_mnras12, kunz_apj12, berlok_apj16}.	

It is assumed, as is customary, that the isotropic part of the pressure tensor --- $P\equiv 2p_\bot/3 + p_\parallel/3$ --- obeys the equation of state for an ideal gas 
$P={\rho k_B T}/{\mu m_H}$,
where $k_B$, $\mu$, and $m_H$ denote the Boltzmann constant, the mean molecular weight, and the atomic mass unit respectively. It then follows that $\alpha T=\alpha_\mu\mu=1$, where 
$\alpha\equiv-({\partial \ln \rho}/{\partial T})_{P,\mu}$ and $\alpha_\mu \equiv({\partial \ln\rho}/{\partial\mu})_{P,T}$ are 
respectively the coefficient of thermal expansion and the coefficient of expansion due to change in the mean molecular weight that is related to the concentration $c$ as follows:
\be
{\mu}=\left[(1-c)\frac{1+Z_1}{\mu_1}+c\frac{1+Z_2}{\mu_2}\right]^{-1}.
\en
Here $Z_i$ and $\mu_i$ ($i\in\{1,2\}$) are respectively the atomic number and 
the molecular weight of the $i$th ion. In our case, $i=1$ for Hydrogen and $i=2$ for Helium.

We are now fully equipped to do linear stability analysis about the equilibrium state. Henceforth an infinitesimal perturbation of a physical quantity, say $f$, about its equilibrium state value has been denoted as $\delta f$. Since the sound speed is much faster than the speed at which the unstable convective modes grows, it is justified to employ the Boussinesq approximation
~\citep{balbus_apj00, balbus_apj01, quataert_apj08, pessah_apj13, himanshu_pla16}, i.e., $\bb{\nabla}\bcdot\delta\mathbf{u}=0$.  
Also one has, due to the solenoidal character of the magnetic field, 
$\bb{\nabla}\bcdot\delta\mathbf{B}=0$. Therefore, we conveniently chose $\delta u_{z}, \delta \omega_{z}, \delta B_{z}$, and $\delta j_{z}$ as the independent variables to work with. $\delta \omega_{z}$ and $\delta j_{z}/4\pi$ respectively denote the $z$-components of the vorticity and the current density fluctuations.

In order to work with the dimensionless equations for the sake of convenience, we define the non-dimensional space and time coordinates as $\bb{x}'\equiv\bb{x}/d$ and $t'\equiv t\nu/d^2$ respectively.
Accordingly, we introduce dimensionless variables $\delta u'_{z}$, $\delta \omega_{z}'$, $\delta B'_z$, $\delta j_{z}'$, $\delta \theta'$, and $\delta c'$ which respectively equal to $\delta u_zd/\lambda$, $\delta \omega_{z} d^2/\lambda$, $\delta B_z/B$, $\delta j_{z} d/B$, $\delta T/\Delta T$, and $\delta c/\Delta c$. Additionally, we make an ansatz that for any (non-dimensional) perturbation of a physical variable, $\delta f'(\bb{x}', t')$=$\sum_{\bb{k}'} \, \hat{f'}(z') \,\exp(i\bb{k}' \bcdot \bb{x}'+\sigma't')\,$ with $\bb{k}'= (k_xd, k_yd, 0)$.  For the sake of brevity without causing any ambiguity, in what follows, we drop all the primes and the hat-symbols.

Now, taking the Laplacian of the $z$-component of momentum equation~(\ref{eq:v})
and the $z$-component of its curl, we arrive at the following equations for
$\delta u_{z}$ and $\delta \omega_{z}$:
\be
\label{eq:pert_W}
&&\Big[\sigma(\partial_z^2-k^2)\textcolor{black}{-}\frac{3}{k^2}(i \sin\!\phi\, k_x+ \cos\!\phi\, \partial_z)^2\times\\
&&( \cos\!\phi\, k^2+i \sin\!\phi\, k_x\partial_z)^2\Big]\delta u_z = -R_Tk^2\delta \theta\nonumber + R_c k^2\delta c\\
&& + Q\f{{\rm P_{\rm r}}}{{\rm P_{\rm m}}}(i \sin\!\phi\, k_x+ \cos\!\phi\, \partial_z)(\partial_z^2-k^2)\delta B_{z}\nonumber\\
&&\textcolor{black}{+}\Big[\frac{3i \sin\!\phi\, k_y}{k^2}(i \sin\!\phi\, k_x+ \cos\!\phi\, \partial_z)^2
(\cos\!\phi\, k^2+i \sin\!\phi\, k_x\partial_z)\Big]\delta \omega_z\,
\nonumber,
\en
{\color{black}
	\be
	\label{eq:pert_Z}
	&&\left[\left(\frac{3\sin^{2}\!\phi\,k_y^2}{k^2}\right)(i\sin\!\phi\,k_x+\cos\!\phi\,\partial_z)^2\right]\delta \omega_z\textcolor{black}{+}\sigma\delta \omega_z= \\
	&&\textcolor{black}{+}Q\f{{\rm P_{\rm r}}}{{\rm P_{\rm m}}}(i\sin\!\phi\,k_x+\cos\!\phi\,\partial_z)\delta j_{z}\nonumber\\
	&&+\Big[\f{3i\sin\!\phi\,k_y}{k^2}(i\sin\!\phi\,k_x+\cos\!\phi\,\partial_z)^2
	\left(\cos\!\phi\,k^2+{i\sin\!\phi\,k_x}\partial_z\right)\Big]\delta u_z\,.
	\nonumber
	\en
}
Here, we have defined the following non-dimensional numbers: Rayleigh number 
(based on $T$) $R_T\equiv{\alpha(\Delta T) gd^3}/{\lambda\nu}$, Rayleigh number 
(based on $c$) $R_c\equiv{\alpha_c(\Delta c) gd^3}/{\lambda\nu}$, Chandrasekhar 
number $Q\equiv{B^2d^2}/{4\pi\rho\nu \eta}$, Prandtl number ${\rm P_{\rm r}}\equiv \nu/\lambda$, and 
magnetic Prandtl number ${\rm P_{\rm m}}\equiv{\nu}/{\eta}$. $\nu$ is the kinematic viscosity arising from the anisotropic part of the pressure tensor~\citep{hollweg_jgr85}. Following a similar procedure for induction equation~(\ref{eq:b2}), we obtain the equations for $\delta B_{z}$ and $\delta j_{z}$, viz.,
{\color{black}
	\be
	&&(\partial_z^2-k^2-{\rm P_{\rm m}}\sigma)\delta B_{z}=-\f{{\rm P_{\rm m}}}{{\rm P_{\rm r}}}(i\sin\!\phi\,k_x+\cos\!\phi\,\partial_z)\delta u_z\,,\\
	&&(\partial_z^2-k^2-{\rm P_{\rm m}}\sigma)\delta j_{z}=-\f{{\rm P_{\rm m}}}{{\rm P_{\rm r}}}(i\sin\!\phi\, k_x+\cos\!\phi\,\partial_z)\delta \omega_z\,.
	\en
}
Similarly, we obtain the equation for the thermal fluctuations from 
equation~(\ref{eq:S}) as
{\color{black}
	\be\label{eq:Temp}
	&&(\cos^{2}\!\phi\,\partial_z^2+2i\sin\!\phi\,\cos\!\phi\,k_x\partial_z-\sin^{2}\!\phi\,k_x^2-{\rm P_{\rm r}} \sigma)\delta \theta \\
	&& =\Sigma\delta u_z+\f{1}{k^2}\left[({A_1}-{A_2})k_xk_y+i{A_5}k_y\partial_z\right]\delta j_{z}+ \nonumber\\
	&& \Big[-\f{A_1}{k^2}\partial_z-\f{A_2}{k^2}k_x^2\partial_z+A_3\partial_z+iA_4k_x+i\f{A_5}{k^2}k_x\partial_z^2\Big]\delta B_{z}\,,\nonumber
	\en
}
where we have used Schwarzschild number $\Sigma \equiv{g\alpha T d}/{c_p\Delta T}-1$.
Lastly, we also have
\be\label{eq:Conc}
&&\left[\Lambda(\cos^{2}\!\phi\,\partial_z^2+2i\sin\!\phi\,\cos\!\phi\,k_x\partial_z-\sin^{2}\!\phi\,k_x^2\textcolor{black}{)}-{\rm P_{\rm r}} \sigma\right] \delta c \\
&& = - \delta u_z+\f{\Lambda}{k^2}\left[({A_1}-{A_2})k_xk_y+i{A_5}k_y\partial_z\right]\delta j_{z}+ \nonumber\\
&& \Lambda \Big[-\f{A_1}{k^2}\partial_z-\f{A_2}{k^2}k_x^2\partial_z+A_3\partial_z+iA_4k_x+i\f{A_5}{k^2}k_x\partial_z^2\Big]\delta B_{z}\,.\nonumber
\en
Here, $\Lambda\equiv D/\lambda$.
In the preceding equations, we have used the notations: $A_1\equiv \cos\!\phi\,,\, A_2\equiv \cos\!\phi\,(\cos^{2}\!\phi-\sin^{2}\!\phi)\,,\,A_3\equiv 2 \sin^{2}\!\phi\, \cos\!\phi\,,\, A_4\equiv \sin\!\phi\, (\sin^{2}\!\phi-\cos^{2}\!\phi)\,, \,A_5\equiv -2\sin\!\phi\, \cos^{2}\!\phi\,$. It may be noted that since we are considering the effect of concentration gradient on our system, equations (\ref{eq:pert_W}) to (\ref{eq:Temp}) can be seen as the extensions of the corresponding analogous equations reported in \citet{himanshu_pla16}. However, equation (\ref{eq:Conc}) is a new equation that arises due to the dynamics of the Helium ion's concentration in the ICM.
\section{Helium Fingers}
The underlying concept of double diffusive convection is most easily intuited in the absence of any magnetic field
and conductivity: Consider ``warm salty over cold fresh" configuration, i.e., 
suppose saline water have the positive vertical gradients of both the temperature $T$ and the salinity $S$, 
but is stably stratified in the sense that the vertical density gradient is negative. In this configuration, 
had the diffusive effects been absent, a blob of fluid {that} is displaced vertically downwards would experience 
an upward buoyancy force because the density of its surrounding is more than the blob's. 
However, under the influence of diffusive processes, the situation becomes different. 
Since the heat diffuses much faster (by about 100 times) than the salt, the downwardly displaced blob 
of fluid can come in the thermal equilibrium with its surrounding while remaining more salty than the immediate surrounding. 
Thus, the blob may become denser than the surrounding and continue to fall under gravity giving way to the salt finger instability. 
This regime of the double diffusive convection is called the finger regime. 
As usual, the most basic techniques of linear stability analysis yield a dispersion relation implicitly relating 
the growth rate of a mode with its corresponding (horizontal) wave number. The inverse of the horizontal wave number associated 
with the fastest growth rate naturally corresponds to the width of the dominant salt fingers observed. 
While near the onset (critical condition) of the instability the convective cells' width is equal to 
the system height, in supercritical condition, one can derive~\citep{stern_tellus60} an expression for 
the width $W$ of the convection cell (salt finger) for the fastest growing mode as,
\begin{equation}\label{e:width}
	W=\left[\frac{g\left(\alpha_T\frac{dT}{dz}-\alpha_S \frac{dS}{dz}\right)}{\nu\lambda}\right]^{-\frac{1}{4}}.
\end{equation}
Here, $\lambda$ is heat diffusion coefficient, $\nu$ is kinematic viscosity, 
$\alpha_T=-({\partial \ln\rho}/{\partial T})_{P,S}$ is thermal expansion coefficient, 
$\alpha_S=({\partial \ln\rho}/{\partial S})_{P,T}$ is expansion coefficient due to salinity. 
It is in the supercritical regime that the convective cells become elongated and can now be 
aptly called salt finger. Note that for the constant temperature and the constant salinity gradients, 
the width $W\propto d^{{1}/{4}}$ where $d$ is the system height.

When the magnetic field and, as is relevant in our case, the magnetic field driven anisotropic transport 
properties comes into effect, the situation is way more complicated both analytically and physically. {Nevertheless, our goal is to investigate the finger regime in the inner region of the ICM within the approximations stated in the earlier sections.} 
This essentially means that we want to study the Helium fingers in the supercritical regime of the HBI modified 
by the concentration gradient. In fact, it can be shown (see Appendix~\ref{sec:appen_A}) that a typical ICM may be in supercritical state. All one has to do is to compare ${d}\ln(T/\mu)/dz$ in the ICM with the corresponding critical value of the gradient at which instability ensues or rather marginal state is attained. This critical value calculated using the convenient conducting stress-free boundary conditions and including the effect of magnetic tension is
	\be
	\left.\f{d}{dz}\ln(T/\mu)\right|_{\rm critical}\approx\left(\f{\pi^2}{\beta d}\right)\,.\label{eq:AppinTex}
	\en
	For an estimate, it may be noted that the critical value of $\mathrm{d}\ln(T/\mu)/\mathrm{dz}\approx 10^{-23}-10^{-25}{\rm cm^{-1}}$ on 
	using $\beta\approx 100-1000$ and $d=10-100$~kpc for a typical inner region of galaxy cluster. 
	However, on using the plot in Figure~5 of \citet{pessah_apj13} based on \citet{vikhlinin_apjl06} 
	and \citet{bulbul_aanda11}, we may take $T_{\rm top}$=7.0 keV, $T_{\rm bottom}$=3.8 keV, 
	$\mu_{\rm top}$=0.83, $\mu_{\rm bottom}$=0.75, and $d$=50 kpc. Thus, from the fiducial profiles 
	of the temperature and the mean-molecular weight, 
	we find that the observed typical value of $\mathrm{d}\ln(T/\mu)/\mathrm{dz}\approx 10^{-23} {\rm cm^{-1}}$. 
	This value evidently can be higher than the critical value. This means that a typical ICM may actually be in a supercritical regime when modelled as a fluid system undergoing double diffusive convection. {This may give us hint that Helium fingers, 
	although not directly observable, may be present in the ICMs.}

In principle, our aforementioned endeavor of investigating the finger regime is straightforward. First, we have to find the mode corresponding to the 
horizontal wave number $k=k_f$ (say) for which growth rate is fastest. Then, the width $W_{\rm He}$ of 
the Helium finger can be determined by putting equal to $2\pi/k_f$. We again want the dependence of the finger width on 
the height of the system. So, we have to chose the expression $g[\alpha (dT/dz)-\alpha_c(dc/dz)]/\nu\lambda$ --- henceforth
denoted as $\xi$ for the sake of convenience --- analogous to the salt finger case. Here, $\alpha_c \equiv({\partial \ln\rho}/{\partial c})_{P,T}$. 
Finally, we ask on which power of $\xi$ the width of the finger depends. 
{In other words, is there a value of $a\in\mathbb{R}$ for which a power-law relation of the following form
	\begin{equation}
		W_{\rm He}=\left[\frac{g\left(\alpha\frac{dT}{dz}-\alpha_c \frac{dc}{dz}\right)}{\nu\lambda}\right]^{a}=\xi^a\,
	\end{equation} 
	is satisfied?} It may be noted that for the constant temperature gradient and the constant concentration gradient, the width $W_{\rm He}\propto d^{-a}$. 

In order to put the idea into practice, we explicitly derive the following relevant dispersion relation by assuming 
the perturbation of the form $\sin(\pi z)e^{i(k_x x+k_y y)+\sigma t}$ and putting $\phi=0$ in 
equations (\ref{eq:pert_W}) to (\ref{eq:Conc}):
\be\label{dispersion}
b_4~\sigma^4+b_3~\sigma^3+b_2~\sigma^2+b_1~\sigma+b_0=0.
\en
Here,
\begin{eqnarray}
	&&b_4=(k^2+\pi^2)~{\rm P_mP_r^2}\,,\\
	&&b_3=(1+\Lambda)~(k^2+\pi^2)~\pi^2{\rm P_mP_r}+(k^2+\pi^2)^2~{\rm P_r^2}\nonumber \\
	&&\hspace{0.6 cm}- 3 k^2~\pi^2~{\rm P_m P_r^2}\,,\\
	&&b_2=\pi^4(k^2+\pi^2)\Lambda{\rm P_m}+(k^2+\pi^2)^2\pi^2(1+\Lambda){\rm P_r}\nonumber \\
	&&\hspace{0.6 cm}-3 k^2\pi^4(1+\Lambda){\rm P_m P_r} - 3 k^2\pi^2(k^2+\pi^2){\rm P_r^2} \nonumber\\
	&&\hspace{0.6 cm}+\pi^2(k^2+\pi^2){\rm P_r^2Q} + k^2{\rm P_mP_r}(R_c + R_T\Sigma\Lambda)\,,\\
	&&b_1=(k^2+\pi^2)^2\pi^4\Lambda- 3 k^2\pi^6\Lambda{\rm P_m} -3 k^2\pi^4(k^2+\pi^2){\rm P_r}(1+\Lambda)\nonumber\\
	&&\hspace{0.6 cm}+(k^2+\pi^2)(1+\Lambda)\pi^4{\rm P_rQ}+k^2~\pi^2~{\rm P_m}~R_T~(1+\Lambda~\Sigma) \nonumber\\
	&&\hspace{0.6 cm}+ k^2~(k^2 + \pi^2)~{\rm P_r}~(R_c+R_T~\Sigma)\,,\\
	&&b_0=-3 k^2\pi^6\Lambda(k^2+\pi^2)+\pi^6~(k^2+\pi^2)\Lambda {\rm Q}+k^2 \pi^2 (k^2+\pi^2) R_c\nonumber\\
	&&\hspace{0.6 cm}+ \frac{k^2 \pi^4 {\rm P_m}(\Lambda R_T-R_c)}{\rm P_r} + k^2\pi^2(k^2+\pi^2)\Lambda R_T\Sigma\,.
\end{eqnarray}
\begin{figure}
	{\includegraphics[width=8.5cm]{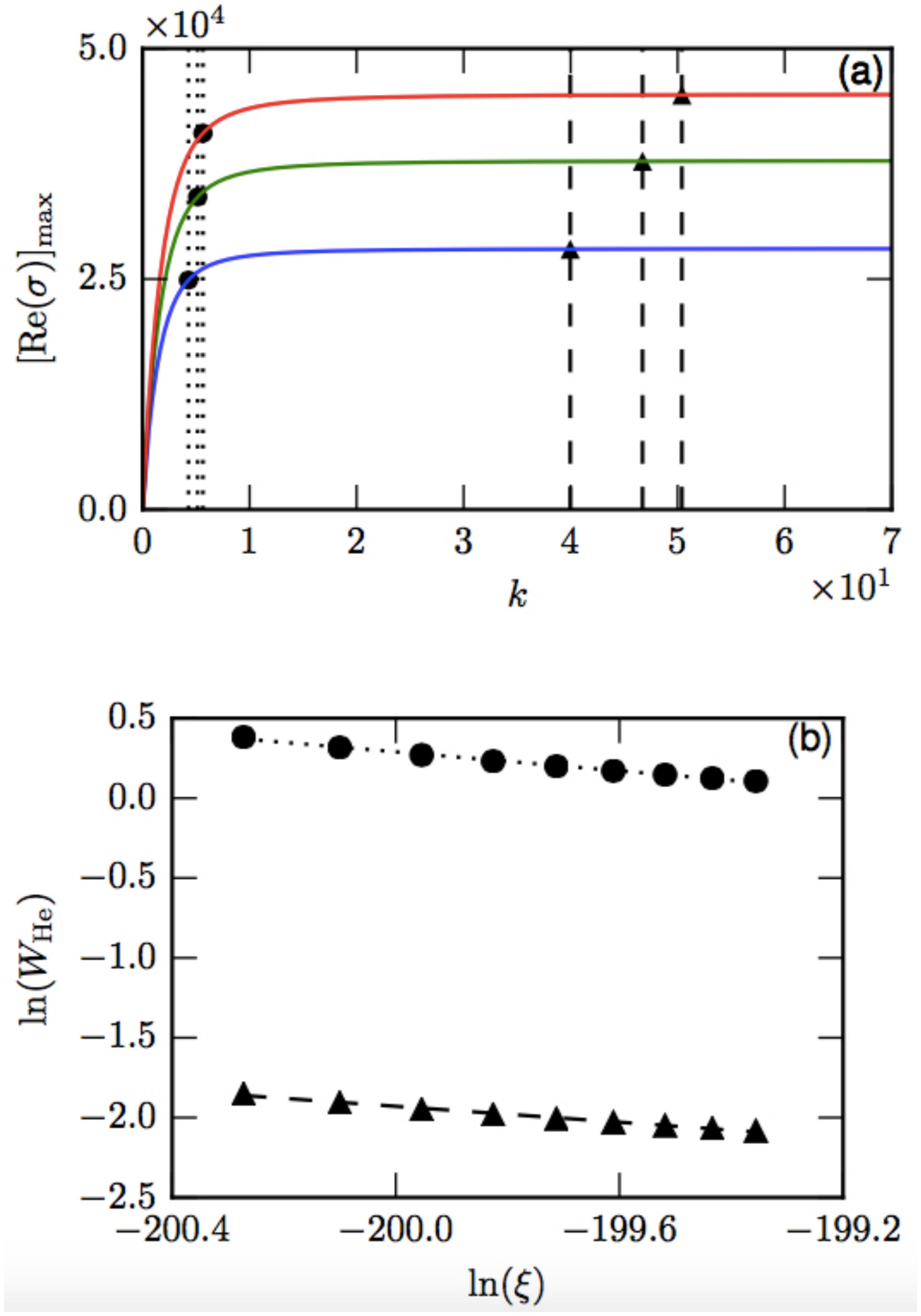}}
	\caption{Width of the Helium fingers follows power law scaling. Upper panel (a) illustrates the largest real part of the growth rate, $[{\rm Re}(\sigma)]_{\rm max}$, as a function of horizontal wavenumber $k$ for three different values of $\xi$ fixed by choosing $\Delta T=-6\times 10^6~K$ (lower blue curve), $\Delta T=-8\times 10^6~K$ (middle green curve), and $\Delta T=-10^7~K$ (upper red curve). The dotted lines and the dashed lines mark the corresponding $[{\rm Re}(\sigma)]_{\rm max}$ for $k$ at which ${d[{\rm Re(\sigma)}]_{\rm max}}/{dk}$ equals $1250$ and $2.5$ respectively. Subsequently,  we choose $9$ different values of $\Delta T$ equispaced between $-10^7$ and $-5\times 10^6~K$, and hence $9$ different values of $\xi$. Using these values of $\xi$, lower panel (b) exhibits two plots of $\ln({ W}_{\rm He}$) vs. $\ln(\xi)$ --- one for ${d[{\rm Re(\sigma)}]_{\rm max}}/{dk}=1250$ (circles) and another for ${d[{\rm Re(\sigma)}]_{\rm max}}/{dk}=2.5$ (triangles). The dotted line of slope $-0.30$ and the dashed line of slope $-0.25$ are the respective best fit lines. In plotting these curves, we have used $\nu=10^{26}~\rm{cm^2 s^{-1}},~ B=2\times 10^{-6}~\rm{G},~\rho=10^{-26}~gm~cm^{-3},~ d= 5\times 10^{22} cm,~\Delta c=-0.2,~c=0.6,~{\rm and}~\eta=100~cm^2~s^{-1}$.}
	\label{f:growth}
\end{figure}
Then we solve for $\sigma$ from this dispersion relation and plot the largest real part of growth rate, 
denoted as $[{\rm Re}(\sigma)]_{\rm max}$, versus $k$ as depicted in Figure~\ref{f:growth}(a). 
We reiterate that we are interested in the inner region of the galaxy cluster, because 
the inner region of the galaxy cluster has {positive} temperature and concentration gradients  which is analogous to what happens in a salt finger system, and hence we choose the system 
parameters appropriately as given in the figure caption. We should also recall that we are dealing with non-dimensionalized quantities for which the unit of distance is $d$ and that of time is $d/\nu^2$. For our case, these units explicitly are $d=5\times10^{22}$~cm and $d/\nu^2=25\times10^{18}$~s respectively. Additionally, it should be noted that our results are concerned with the local modes in fluid limit as they should since the typical Knudsen number (ratio of ionic mean free path to the scale height) in the inner region of cool-core cluster is of the order of $10^{-3}$~\citep{kunz_mnras11, pessah_apj13}.

We note that $[{\rm Re}(\sigma)]_{\rm max}$ increases monotonically with $k$ and $[{\rm Re}(\sigma)]_{\rm max}$ 
corresponds to the wave number $k=k_f=\infty$. This is consistent with the fact that the the lowest mode first 
go unstable at $k_f$ as detailed in the preceding paragraphs. This is not only inconvenient to work with but 
also is not physically allowed within the scope of our calculations: in our analysis, it doesn't make 
sense to choose a $k$ that is larger than the inverse of the mean free path of the ions in the ICM, 
otherwise we are no longer in the fluid limit for the system under consideration. In order to bypass this issue, 
we use the fact that the curve of $[{\rm Re}(\sigma)]_{\rm max}$ vs. $k$ ($\xi$ fixed) is monotonic and 
appear to be saturating with $k$. We choose that $k_f$ as $k$ for which ${d [{\rm Re}(\sigma)]_{\rm max}}/{dk}$ is 
small, say, $1250$. This value is chosen so that at this value the curve enters into the region 
where ${d [{\rm Re}(\sigma)]_{\rm max}}/{dk}$ no longer changes appreciably. 
Of course, strictly speaking this $k_f$ does not correspond to the fastest growing perturbation. 
In Figure~\ref{f:growth}(a), we have plotted $[{\rm Re}(\sigma)]_{\rm max}$ with the horizontal wave 
number $k$ for different values of $\xi$ by varying the temperature difference between the two plates confining the plasma. 
We then choose $k_f$ for each curve and thence find the corresponding Helium finger widths, $W_{\rm He}$. 
Consequently, $\ln(W_{\rm He})$ vs. $\ln(\xi)$ can be plotted (see Figure~\ref{f:growth}(b)) and 
from the slope of the resulting curve, we can get the value of $a$. We, intriguingly, have found $a\approx -0.30$, 
very close to that is obtained for salt fingers (see equation \ref{e:width}). 

In passing, we must make it clear that, in principle, one should have strictly worked with the fastest growing mode but that 
is not appropriate as discussed in the preceding paragraph. Instead we have used somewhat ad hoc modes and investigated its 
growth. 
This means that the power law exponent $a$  is prone to modification in a realistic ICM, e.g., 
{when we select ${d [{\rm Re}(\sigma)]_{\rm max}}/{dk}= 2.5$ which corresponds to $k_f$s greater than the ones obtained for ${d [{\rm Re}(\sigma)]_{\rm max}}/{dk}= 1250$, we get $a=-0.25$ that is even more close (within curve-fitting errors) to $-1/4$}. Nevertheless, numerical simulations~(see Figure 7 in \citet{berlok_apj16}) 
validates the existence of thin Helium fingers. These fingers are observed en route to the turbulence state of the ICM.

\section{Discussions and Conclusions} 
\label{sec:DaC}

In this paper, we have applied the formalism usually employed for studying salt finger instability to investigate 
analogous convective instabilities in the inhomogeneous ICM. This approach goes beyond the standard
linear mode analysis and has been carried out (however see
\citealt{latter_mnras12}) by explicitly accounting for the boundary conditions commonly employed in the numerical simulations. 
Our work facilitates analytical understanding of some hitherto unexplored aspects of the dynamics of these instabilities. 
Moreover, in principle, this formalism is generalizable to address more realistic physical settings 
that, e.g., includes the effects of cosmic-rays \citep{chandran_apj06}, radiative cooling
\citep{balbus_apjl10, latter_mnras12}, shear~\citep{ren_pop11}, rotation \citep{nipoti_apj14}, etc.

	As far as the stability analyses of such inhomogeneous ICM modelled as weakly 
	collisional dilute binary mixture of ions are 
	concerned~\citep{pessah_apj13, berlok_apj15, berlok_apj16}, for the first time we have an exact analytical expression 
	for the stability criterion (equation (\ref{eq:AppinTex})) that takes into account the effect of the magnetic tension. 
	This analytical expression till now could not be arrived at using more widely employed and less cumbersome WKB-type perturbative 
	methods~\citep{balbus_apj01, quataert_apj08, pessah_apj13, berlok_apj15}. However, it must be acknowledged that the criterion in principle could be obtained numerically and that the magnetic tension has quantitative effect on the stability criterion can easily be seen within the WKB-formalism.
	{In any case}, using this result, we have deduced that many cool-core galaxy clusters may be in the supercritical regime and 
	hence there is the possibility of existence of Helium fingers in the inner region of the ICMs. Moreover, analogous to the 
	finger regime of double diffusive instabilities in the absence of the magnetic field, the Helium fingers' width 
	goes approximately as one-fourth power of the separation between the plates holding the fluid in the presence of
	vertical magnetic field --- this is the main result of the paper. This intuitively makes sense simply because, as the Helium fingers are 
developing vertically downwards, the motion is along the magnetic field and hence the magnetic field is not presenting 
any obstruction to it via the Lorenz force. We remark that, as discussed earlier, even though the exact value of the exponent may not be 
robust, the power-law nature may still be. {However, it is encouraging to note that the power law exponent 
	$a$ approaches $-1/4$ as one chooses larger value for $k_f$}; this fact is compatible with the aforementioned intuition.

This, however, may not be the case if the magnetic field is taken to be horizontal. 
We know that under the assumption of homogeneous ICM this case is more relevant in the outer region 
of ICM where the gradient of the temperature is typically negative giving rise to the MTI. 
Interestingly, there composition gradient could also be negative. This is reminiscent of diffusive 
regime~\citep{kundu_book12} of double diffusive instabilities leading to oscillatory instability in 
the corresponding saline water. Such oscillatory instability eventually results in the formation of a 
number of horizontal convecting layers. Analogously, thus, we expect similar horizontal convecting 
layers in the outer region of the ICM but of course the effect of the magnetic field could be significant. 
This issue of subtle interplay between the magnetic field and the flow patterns is worth pursuing as 
a future research direction.

Before we end, it must be admitted that the idealized setup of inhomogeneous plasma between two fixed plates 
along with the simplified boundary conditions, we have worked with in this paper can not be the complete 
picture of the ICM. Nevertheless, we have seen that it allows us to shed light on some aspects of double 
diffusive convection processes --- generalization of the HBI and the MTI in the presence of the composition 
gradient --- that play important role in the dynamics of the ICM. Some of the more positives of this particularly 
simple yet insightful setup could be as follows: Firstly, this approach provides a practical platform to 
make connections with numerical simulations because the boundary conditions usually
adopted resemble the ones employed herein. Secondly, it identifies the critical value of the relevant gradients 
for the onset of the instabilities. Specifically, it can account for the effects of the magnetic tension on the 
stability criteria for both the HBI and the MTI. Thirdly, it provides an ideal platform for carrying out weakly
nonlinear stability analysis for the convective instabilities. However, such {a challenging} analysis is infested 
with subtle caveats and technicalities. Last but not the least, such analyses could enable invention of low 
dimensional models like the Lorenz model for Rayleigh--B\'enard convection \citep*{lorenz_jas63,
	chen_csf06} and magnetic Rayleigh--B\'enard convection \citep{zierep_tam03}, capable of rendering further 
insights into the chaotic (turbulent) state of the ICM. As far as the choice of the boundary conditions are 
concerned, desire for analytical simplicity directed us to adopt conducting
stress-free boundary conditions. They resemble those conveniently employed
in numerical simulations~\citep{berlok_apj16,berlok_apj16b}. Of course, one could have worked with non-conducting
stress-free, conducting rigid, non-conducting rigid, or some other possible boundary conditions but then it 
is not always possible to solve the corresponding problem analytically and even in the
linear regime numerical techniques may become indispensible. Although, a particular choice 
of the boundary conditions is unlikely to have a drastic impact in 
the stability criterion within the bulk of the plasma, it must be realised that it may 
alter the detailed expression for the stability criterion.

\section*{Acknowledgements} We are thankful to Thomas Berlok, Martin E. Pessah, Shailendra K. Rathor, and Manohar K. Sharma for useful discussions. Part of this work was done during the visit  of S.~C. at TIFR Centre for Interdisciplinary Sciences, Hyderabad, India as a visiting faculty. S.~C.~acknowledges the financial support through the INSPIRE faculty award (DST/INSPIRE/04/2013/000365) conferred 
by the Indian National Science Academy (INSA) and the Department of Science and Technology (DST), India.
% Appendixes {{{1
% ---------------

\appendix

% ---------------
\section{Criterion for marginal state in heat- and particle-flux-driven buoyancy instability}
\label{sec:appen_A}
In what follows, we exclusively adopt the reflective, stress-free, and
	perfectly conducting boundaries mathematically given by
	\be
	&&\delta u_z(0)=\delta u_z(1)=0\,,\label{W1}\\
	&&\partial_z^2\delta u_z(0)=\partial_z^2\delta u_z(1)=0\,,\label{D2W}\\
	&&\partial_z\delta \omega_z(0)=\partial_z\delta \omega_z(1)=0\,\label{DZ}, \\
	&&\delta B_{z}(0)=\delta B_{z}(1)=0\,,\label{K}\\
	&&\partial_z\delta j_{z}(0)=\partial_z\delta j_{z}(1)=0\,,\label{DX} \\
	&&\delta \theta(0)=\delta \theta(1)=0\,\label{theta1} \,,\\
	&&\delta c(0)=\delta c(1)=0\,.\label{conc1}
	\en
	Equation (\ref{W1}) means that the normal component of the velocity is
	zero on the boundary surfaces. Equation (\ref{D2W}) and 
	equation(\ref{DZ}) imply stress-free surfaces, whereas equation (\ref{K}) and equation (\ref{DX}) 
	model perfectly conducting boundaries. Equation (\ref{theta1}) and equation (\ref{conc1}) force the boundary
	surfaces to be at a constant temperature and concentration respectively. While facilitating analytically tractable calculations, this set of boundary conditions is also straightforwardly implementable in the numerical simulations.

Although in general $\phi\ne0$, it is convenient to consider the case in which the magnetic field is
	along the $z$-direction, i.e., $\phi=0$, that is known to be prone
	to the heat- and particle-flux-driven buoyancy instability~\citep{pessah_apj13} which itself is 
	the generalization of HBI in the presence of Helium concentration. This analytically tractable case is 
	rich enough to make us appreciate the possible physical origin of the Helium fingers in the ICM. Hence, 
	from now on we shall {focus on} this case.

In the state of marginal stability $(\sigma=0)$, the system of 
	equation~(\ref{eq:pert_W}) and equation (\ref{eq:pert_Z}) are combined to give
	\be
	\left[\textcolor{black}{-}3k^2\partial_z^4 + \left(\Sigma R_T+\f{R_c}{\Lambda} \right)k^2 \right](\partial_z^2-k^2)\delta u_z=\nonumber \\
	-\left[(R_T-R_c)\f{{\rm P_{\rm m}}}{{\rm P_{\rm r}}}k^2+Q(\partial_z^2-k^2)\partial_z^2\right]\partial_z^2\delta u_z\,.\label{eq:Wmcfull}
	\en
	Now, it is known that the dimensionless parameters $Q$ and ${\rm P_{\rm m}}$ have 
	extremely large values in the ICM~ \citep{carilli_araa02, peterson_pr06,himanshu_pla16}. 
	Thus, in this limit of interest, we may choose $Q,{\rm P_{\rm m}}\rightarrow\infty$, causing
	equation (\ref{eq:Wmcfull}) to reduce to
	\be
	\left[(R_T-R_c)\f{{\rm P_{\rm m}}}{{\rm P_{\rm r}}}k^2+Q(\partial_z^2-k^2)\partial_z^2\right]\partial_z^2\delta u_z=0\,.\label{eq:Wmc}
	\en
	Now, the application of the chosen boundary conditions allows us to 
	conclude~(details are analogous to the calculations in \citet{himanshu_pla16}) that
	\be
	\delta u_z=A\sin n\pi z\,,
	\en
	($A$ being constant), along with
	\be
	&&R_T-R_c=-n^2\pi^2 \left[\f{n^2\pi^2+k^2}{k^2}\right]Q\f{{\rm P_{\rm r}}}{{\rm P_{\rm m}}}\,.\label{eq:mRscv}\\
	\Rightarrow&&\f{d}{dz}\ln(T/\mu)=n^2\pi^2 \left[\f{n^2\pi^2+k^2}{k^2}\right]\left(\f{1}{\beta d}\right)\,.\label{eq:mRscv1}
	\en
	Since the R.H.S. of  equation~({\ref{eq:mRscv1}}) is non-negative, the marginal state can exist 
	when (i) $\Delta T$ is negative {i.e.,} top plate is hotter and $\Delta\mu>0$, 
	or (ii) $\Delta T$ is positive (bottom plate hotter) and $\Delta\mu>0$ but $|\Delta \ln T|<|\Delta\ln\mu|$, 
	or (iii) $\Delta \mu$ and $\Delta T$ are both negative, {i.e.,} top plate is hotter and at higher concentration,
	such that $|\Delta \ln T|>|\Delta\ln\mu|$. The last case may be satisfied (in accordance with the Helium sedimentation model) in the inner region of the ICM in 
	cool-core galaxy cluster. Here, the onset of instability happens for the lowest mode ($n=1$) for the maximum 
	value of $k=\infty$ at which $({\pi^2+k^2})/{k^2}$ 
	attains the minimum value $1$. Therefore, the criterion for the onset of instability is
	\be
	\left.\f{d}{dz}\ln(T/\mu)\right|_{\rm critical}\approx\left(\f{\pi^2}{\beta d}\right)\,,\label{eq:critical}
	\en
	One should note that the right hand side of equation (\ref{eq:critical}) is  quite small for reasonable values of $\beta$ in the ICM and hence onset of instability is quite probable for the typical values of the temperature and concentration gradients as has been argued in the main text of this paper.
	The `equal to' sign has not been {used in} this stability criterion because it must be kept in mind that in order to remain 
	within the fluid limit, $k$ can at most be of the order of inverse of the mean free path of the ions. Similar objection can be raised 
	against the usage of $n=1$ as it isn't compatible with the local approximation; this however is not a very serious concern since choosing any larger value of $n$
	only modifies the constant multiplicative numerical factor in the R.H.S. of equation~(\ref{eq:critical}). Most important aspect
	of the mathematical expression of the criterion is that it takes into account the effect of magnetic field strength (or equivalently the timescale of Alfv\'en velocity). Such an analytical expression is missing in the work done by \citet{pessah_apj13, berlok_apj15, berlok_apj16} and hence equation 
	(\ref{eq:critical}) is a new result. It explicitly shows that the critical limit of the gradient of $(T/\mu)$ increases proportional to the square of the strength of the background magnetic field.
%
%%%%%%%%%%%%%%%%%%%%%%%%%%%%%%%%%%%%%%%%%%%%%%%%%%%%%%%%%%%%%%%%%%%
%\section*{References}
\bibliographystyle{mn2e}
\bibliography{Sadhukhan_etal_bibliography}

\end{document}